\documentclass[useAMS,usenatbib,a4paper]{mn2e}

\usepackage{graphics}
\usepackage{graphicx}
\usepackage{times}
\usepackage{amsmath}

\usepackage[usenames]{color}

\newcommand{\apj}{ApJ}
\newcommand{\aap}{A\&A}
\newcommand{\apjl}{ApJ}
\newcommand{\apjs}{ApJS}
\newcommand{\aj}{AJ}
\newcommand{\pasp}{PASP}
\newcommand{\mnras}{MNRAS}

\title[Interferometric radii of $\theta$ Cyg and 16 Cyg]
{Interferometric radii of bright {\it Kepler} stars with the CHARA Array: $\theta$~Cygni and 16~Cygni~A~and~B}

\author[T. R. White et al.]
{T. R. White$^{1,2,3}$\thanks{E-mail: t.white@physics.usyd.edu.au}, D. Huber$^{1,4,9}$, V. Maestro$^{1}$, 
T. R. Bedding$^{1,3}$, M. J. Ireland$^{1,2,5}$, 
\newauthor F. Baron$^{6}$, T. S. Boyajian$^{7,8}$, X. Che$^{6}$,
J. D. Monnier$^{6}$, B. J. S. Pope$^{1}$,  
\newauthor R. M. Roettenbacher$^{6}$, D. Stello$^{1,3}$, P. G. Tuthill$^{1}$, 
C. D. Farrington$^{7}$, 
\newauthor P. J. Goldfinger$^{7}$, H. A. McAlister$^{7}$, G. H. Schaefer$^{7}$, J. Sturmann$^{7}$, 
\newauthor L. Sturmann$^{7}$, T. A. ten Brummelaar$^{7}$ and N. H. Turner$^{7}$\\
$^{1}$Sydney Institute for Astronomy (SIfA), School of Physics, University of 
Sydney, NSW 2006, Australia\\
$^{2}$Australian Astronomical Observatory, PO Box 296, Epping, NSW 1710, Australia\\
$^{3}$Stellar Astrophysics Centre, Department of Physics and Astronomy, Aarhus University, Ny Munkegade 120, DK-8000 Aarhus C, Denmark\\
$^{4}$NASA Ames Research Center, MS 244-30, Moffett Field, CA 94035, USA\\
$^{5}$Department of Physics and Astronomy, Macquarie University, NSW 2109, Australia\\
$^{6}$University of Michigan Astronomy Department, 941 Dennison Bldg, Ann Arbor, MI 48109-1090, USA\\
$^{7}$Center for High Angular Resolution Astronomy, Georgia State University, PO Box 3969, Atlanta, GA 30302, USA\\
$^{8}$Yale University, Department of Astronomy, 260 Whitney Ave, New Haven, CT 06520\\
$^{9}$NASA Postdoctoral Program Fellow} 
\begin{document}

\date{Accepted 2013 May 7.  Received 2013 May 3; in original form 2013 March 31}

\pagerange{\pageref{firstpage}--\pageref{lastpage}} \pubyear{2013}

\maketitle

\label{firstpage}

\begin{abstract}
We present the results of long-baseline optical interferometry observations using the Precision Astronomical 
Visual Observations (PAVO) beam combiner at the Center for High Angular Resolution Astronomy (CHARA) Array to measure the 
angular sizes of three bright {\it Kepler} stars: $\theta$~Cygni, and both components of the binary system 16~Cygni.
Supporting infrared observations were made with the Michigan Infrared Combiner (MIRC) and Classic beam combiner,
also at the CHARA Array. We find limb-darkened angular diameters of $0.753\pm0.009$\,mas for $\theta$~Cyg, 
$0.539\pm0.007$\,mas for 16~Cyg~A and $0.490\pm0.006$\,mas for 16~Cyg~B. The {\it Kepler Mission} has observed these 
stars with outstanding photometric precision, revealing the presence of solar-like oscillations. Due to the brightness
of these stars the oscillations have exceptional signal-to-noise, allowing for detailed study through asteroseismology,
and are well constrained by other observations. We have combined our interferometric diameters with Hipparcos parallaxes,
spectrophotometric bolometric fluxes and the asteroseismic large frequency separation to measure linear radii
($\theta$\,Cyg:~1.48$\pm$0.02\,R$_\odot$, 16\,Cyg\,A:~1.22$\pm$0.02\,R$_\odot$, 16\,Cyg\,B:~1.12$\pm$0.02\,R$_\odot$),
effective temperatures ($\theta$\,Cyg:~6749$\pm$44\,K, 16\,Cyg\,A:~5839$\pm$42\,K, 16\,Cyg\,B:~5809$\pm$39\,K),
and masses ($\theta$\,Cyg:~1.37$\pm$0.04\,M$_\odot$, 16\,Cyg\,A:~1.07$\pm$0.05\,M$_\odot$, 16\,Cyg\,B:~1.05$\pm$0.04\,M$_\odot$)
for each star with very little model dependence. The measurements presented here will 
provide strong constraints for future stellar modelling efforts.
\end{abstract}


\section{Introduction}
Progress in understanding stellar structure and evolution is driven by ever-more precise measurements
of fundamental properties such as stellar temperature, radius and mass \citep[see, e.g.][]{Demarque86, Monteiro96, 
Deheuvels11, Trampedach11, Piau11}. Unfortunately, many methods to determine such properties are indirect and, being model-dependent 
themselves, are of little use in improving stellar models. We therefore look towards methods that either themselves, or 
in combination with other methods, have little model dependence.

One such method is asteroseismology, the study of stellar oscillations. Stars like the Sun exhibit many 
convectively-excited oscillation modes whose properties depend on the structure of the star. This allows stellar 
parameters such as mean stellar density and surface gravity to be accurately determined with little model dependence 
\citep[see, e.g.][]{Brown94,C-D04,Aerts10}. 

Another method is long-baseline optical interferometry (LBOI), which can be used to measure the angular sizes of stars. 
Combining with a parallax measurement yields the linear radius, while combining with the bolometric flux provides a direct 
measurement of effective temperature \citep[see, e.g.][]{Code76, Baines09, Boyajian09, Boyajian12a, Boyajian12b, Creevey12b}.

The combination of asteroseismology and interferometry therefore allows us to determine mass, radius and 
temperature with very little model dependence. While the potential value of this has long been recognised 
\citep{Cunha07}, until recently the inherent difficulties in these methods had limited their application in cool stars 
to a few bright objects \citep{North07, Bruntt10, Bazot11}. Progress in instrumentation, such as the Precision 
Astronomical Visible Observations (PAVO) beam combiner \citep{Ireland08} at the Center for High Angular Resolution 
Astronomy (CHARA) Array \citep{tenBrummelaar05}, has pushed the sensitivity limits of LBOI.
Meanwhile asteroseismology has entered a `golden age', thanks to data from the space telescopes {\it CoRoT} 
\citep{Michel08} and {\it Kepler} \citep{Koch10, Gilliland10, Chaplin11a}. 

\citet{Huber12b} recently presented interferometric observations using the PAVO beam combiner at CHARA of F, G and K 
stars spanning from the main sequence to the red clump in which solar-like oscillations have been detected by 
{\it Kepler} or {\it CoRoT}. In this paper we present results from the same instrument, of three bright 
{\it Kepler} targets, $\theta$~Cygni and 16~Cygni~A~\&~B. These targets present an excellent opportunity to 
combine the remarkably precise, high signal-to-noise asteroseismology data from the {\it Kepler Mission} with precise 
constraints from interferometry and other methods, for strong tests of stellar models.

\section{Targets}

\subsection{$\theta$ Cyg}
The F4V star $\theta$ Cygni (13~Cyg, HR~7469, HD~185395, KIC~11918630), magnitude $V=4.48$, 
is the brightest star being observed by {\it Kepler}. Stellar parameters available from the literature are given in 
Table~\ref{tab0}. Thus far {\it Kepler} has observed it in 2010 June--September ({\it Kepler} Quarter 6), 
2011 January--March (Q8) and 2012 January--October (Q12 -- Q14).

A close companion of $V\sim$12 mag has been identified as an M~dwarf, with an estimated mass of 0.35\,M$_\odot$,
separated from the primary by $\sim$2 arcsec, a projected separation of 46\,AU \citep{Desort09}. 
Although it has been detected several times since 1889 \citep{Mason01}, the orbit is still very incomplete.

\citet{Desort09} undertook a radial velocity study of $\theta$ Cyg, finding a 150\,d quasi-periodic variation. The
origin of this variation is still not satisfactorily explained -- the presence of one or two planets does not adequately 
explain all the observations and stellar variation of this period is unknown in stars of this type.

The limb-darkened angular diameter of $\theta$ Cyg has previously been estimated as 
$\theta_\mathrm{LD}$\,=0.760\,$\pm$\,0.021\,mas from spectral energy distribution fitting to photometric observations
by \citet{vanBelle08}. In 2007 and 2008, \citet{Boyajian12a} made interferometric observations with the CHARA Classic beam 
combiner in $K^{\prime}$-band ($\lambda_0=2.14\mu$m). They measured a larger diameter, 
$\theta_\mathrm{LD}$\,=0.861\,$\pm$\,0.015\,mas. \citet{Ligi12}, using the VEGA beam combiner at CHARA found
$\theta_\mathrm{LD}$\,=0.760\,$\pm$\,0.003\,mas, in agreement with \citet{vanBelle08}. \citet{Ligi12} also reported 
excessive scatter in their measurements and speculated on diameter variability or the existence of a new close 
companion, possibly related to the quasi-periodic variability seen in radial velocity by \citet{Desort09}.

The location of $\theta$ Cyg in the HR diagram places it amongst $\gamma$~Dor pulsators. Analysis of 
Q6 {\it Kepler} data by \citet{Guzik11} did not reveal $\gamma$~Dor pulsations, but clear evidence of
solar-like oscillations was seen in the power spectrum between 1200 and 2500~$\mu$Hz. The characteristic large
frequency separation between modes of the same spherical degree, $\Delta\nu$, is 84.0$\pm$0.2~$\mu$Hz. The
oscillation modes are significantly damped resulting in large linewidths in the power spectrum, which is typical 
of F stars \citep{Chaplin09,Baudin11,Appourchaux12a, Corsaro12}.

\begin{table}
 \centering
 \begin{minipage}{140mm}
  \caption{Properties of Target Stars from Available Literature}
  \begin{tabular}{@{}lccc@{}}
  \hline
    & $\theta$ Cyg & 16 Cyg A & 16 Cyg B\\
 \hline
Spectral Type & F4V & G1.5V & G3V\\
$V$ mag & 4.48 & 5.96 & 6.2\\
$T_\mathrm{eff}$ (K) & 6745$\pm$150\footnote{\citet{Erspamer03}, high-resolution spectroscopy} & 5825$\pm$50\footnote{\citet{Ramirez09}, high-resolution spectroscopy} & 5750$\pm$50$^{\displaystyle b}$\\
log $g$ & 4.2$\pm$0.2$^{\displaystyle a}$ & 4.33$\pm$0.07$^{\displaystyle b}$ & 4.34$\pm$0.07$^{\displaystyle b}$\\
$\mathrm{[Fe/H]}$ & $-$0.03$^{\displaystyle a}$ & 0.096$\pm$0.026$^{\displaystyle b}$ & 0.052$\pm$0.021$^{\displaystyle b}$\\
Parallax (mas) & 54.54$\pm$0.15\footnote{\citet{vanLeeuwen07}, revised Hipparcos parallax} & 47.44$\pm$0.27$^{\displaystyle c}$ & 47.14$\pm$0.27$^{\displaystyle c}$\\
Distance (pc) & 18.33$\pm$0.05 & 21.08$\pm$0.12 & 21.21$\pm$0.12 \\
$F_\mathrm{bol}$ (pW.m$^{-2}$) & 392.0$\pm$0.4\footnote{\citet{Boyajian13}, spectrophotometry} & 112.5$\pm$0.2$^{\displaystyle d}$ & 91.08$\pm$0.14$^{\displaystyle d}$ \\
Luminosity (L$_\odot$) & 4.11$\pm$0.02 & 1.56$\pm$0.02 & 1.28$\pm$0.01 \\
Mass (M$_\odot$) & $1.39^{+0.02}_{-0.01}$\footnote{\citet{Casagrande11}, fit to BaSTI isochrones} & 1.11$\pm$0.02\footnote{\citet{Metcalfe12}, asteroseismology} & 1.07$\pm$0.02$^{\displaystyle f}$\\
Radius (R$_\odot$) & ... & 1.243$\pm$0.008$^{\displaystyle f}$ & 1.127$\pm$0.007$^{\displaystyle f}$\\
Age (Gyr) & $1.13^{+0.17}_{-0.21}$$^{\displaystyle e}$ & 6.9$\pm$0.3$^{\displaystyle f}$ & 6.7$\pm$0.4$^{\displaystyle f}$\\
\hline
\label{tab0}
\end{tabular}
\end{minipage}
\end{table}

\subsection{16 Cyg A \& B}\label{16Cyglit}
Our other targets are the solar analogues 16~Cygni~A (HR~7503, HD~186408, KIC~12069424) and B (HR~7504, HD~186427, 
KIC~12069449). Properties of the stars from the literature are listed in Table~\ref{tab0}. {\it Kepler} 
observations between June 2010 and October 2012 (Q6 -- Q14) are currently available.

The separation of the A and B components on the sky is 39.56 arcsec, which enables them to be observed independently 
by both {\it Kepler} and PAVO. They also have a distant M~dwarf companion, about 10 mag fainter, in a hierarchical 
triple system \citep{Turner01, Patience02}. There are, however, no dynamical constraints on their masses due to the 
long orbital period, estimated at over 18,000~years \citep{Hauser99}. Additionally, 16~Cyg~B is known to have a planet 
with a mass of $\sim$ 1.5 M$_\mathrm{J}$ in an eccentric 800 day orbit \citep{Cochran97}.

Interferometric observations of 16~Cyg~A~and~B with the CHARA Classic beam combiner have been presented previously. Observing
in $K^{\prime}$-band, \citet{Baines08} measured a limb-darkened angular diameter for 16~Cyg~B of 
$\theta_\mathrm{LD}$\,=0.426\,$\pm$\,0.056\,mas, although their estimate from a spectral energy distribution fit was somewhat
larger ($\theta_\mathrm{LD}$\,=0.494\,$\pm$\,0.019\,mas). More recently, \citet{Boyajian13} measured both stars with Classic in
$H$-band ($\lambda_0=1.65\mu$m), finding $\theta_\mathrm{LD}$\,=0.554\,$\pm$\,0.011\,mas and 
$\theta_\mathrm{LD}$\,=0.513\,$\pm$\,0.012\,mas for the A and B components, respectively.

{\it Kepler} observations clearly show solar-like oscillations in both stars, with large separations, $\Delta\nu$, of 
103.4~$\mu$Hz and 117.0~$\mu$Hz, respectively \citep{Metcalfe12}. Asteroseismic 
modelling was performed by \citet{Metcalfe12} using several different methods. The values they obtained for mass, radius 
and age are given in Table~\ref{tab0}. Promisingly, although both stars were modelled independently, the models find a 
common age and initial composition, which is to be expected in a binary system. However, inspecting the individual 
results of each model method reveals two families of solutions. Several models favour a radius of $1.24\,\mathrm{R_\odot}$ 
for 16~Cyg~A and $1.12\,\mathrm{R_\odot}$ for 16~Cyg~B, while others favour a larger radii around $1.26\,\mathrm{R_\odot}$
and $1.14\,\mathrm{R_\odot}$, respectively. For comparison, the estimated systematic uncertainties in radius are 
$0.008\,\mathrm{R_\odot}$ and $0.007\,\mathrm{R_\odot}$, respectively.

\section{Observations}\label{obs}

\begin{table}
 \centering
 \begin{minipage}{140mm}
  \caption{Log of PAVO interferometric observations.}
  \begin{tabular}{@{}lcrcc@{}}
  \hline
   UT Date & Baseline\footnote{The baselines used have the following lengths: \\W1W2, 107.92\,m; E2W2, 156.27\,m;
   S2W2, 177.45\,m; S1W2, 210.97\,m;\\S2E2, 248.13\,m; E2W1, 251.34\,m.} & Target & No. Scans & Calibrators\footnote{Refer to Table \ref{tab2} for details of the calibrators used.}\\
 \hline
2010 July 20   & S2E2 & 16 Cyg A     & 1 & c\\
2011 May 27    & E2W2 & $\theta$ Cyg & 3 & ij\\
2011 May 28    & E2W2 & $\theta$ Cyg & 2 & b\\
2011 July 4    & S1W2 & 16 Cyg A     & 3 & bh\\
               &      & 16 Cyg B     & 3 & bh\\
2011 September 9 & S2W2 & 16 Cyg A    & 3 & be\\
               &      & 16 Cyg B     & 3 & be\\
2012 August 4  & S1W2 & 16 Cyg A     & 3 & aeg\\
               &      & 16 Cyg B     & 3 & aeg\\
               &      & $\theta$ Cyg & 3 & aegi\\
2012 August 6  & E2W2 & $\theta$ Cyg & 4 & egi\\
2012 August 8  & S1W2 & 16 Cyg A     & 3 & afi\\
               &      & 16 Cyg B     & 3 & afi\\
               &      & $\theta$ Cyg & 3 & afgi\\
2012 August 9  & S2E2 & 16 Cyg A     & 3 & fgi\\
               &      & 16 Cyg B     & 3 & fgi\\
2012 August 10 & S2W2 & 16 Cyg A     & 2 & bdi\\
               &      & 16 Cyg B     & 2 & dfi\\
               &      & $\theta$ Cyg & 3 & fgi\\
2012 August 11 & W1W2 & $\theta$ Cyg & 1 & i\\
2012 August 12 & E2W1 & 16 Cyg A     & 3 & fik\\
               &      & 16 Cyg B     & 3 & fik\\
2012 August 14 & S2E2 & 16 Cyg A     & 3 & fgi\\
               &      & 16 Cyg B     & 3 & fgi\\
\hline
\label{tab1}
\end{tabular}
\end{minipage}
\end{table}

\begin{table}
 \centering
 \begin{minipage}{140mm}
  \caption{Calibrators used for interferometric observations.}
  \begin{tabular}{@{}lccrccc@{}}
  \hline
   HD & Sp. Type & $V$ & $V-K$ & $E(B-V)$ & $\theta_{V-K}$ & ID \\
 \hline
176626 &   A2V  & 6.85 &  $0.084$ & 0.026 & 0.146 & a\\
177003 & B2.5IV & 5.38 & $-0.524$ & 0.023 & 0.198 & b\\
179483 &   A2V  & 7.21 &  $0.316$ & 0.028 & 0.144 & c\\
180681 &   A0V  & 7.50 &  $0.112$ & 0.031 & 0.111 & d\\
181960 &   A1V  & 6.23 &  $0.121$ & 0.042 & 0.200 & e\\
183142 &   B8V  & 7.07 & $-0.462$ & 0.060 & 0.093 & f\\
184787 &   A0V  & 6.68 &  $0.034$ & 0.017 & 0.154 & g\\
188252 &  B2III & 5.90 & $-0.461$ & 0.047 & 0.156 & h\\
188665\footnote{For this star we instead use the calibrated diameter, \\$\theta_\mathrm{UD}=0.274\pm0.008$ (see text).} &   B5V  & 5.14 & $-0.384$ & 0.035 & 0.240 & i\\
189296 &   A4V  & 6.16 &  $0.250$ & 0.033 & 0.225 & j\\
190025 &   B5V  & 7.55 & $-0.230$ & 0.157 & 0.084 & k\\ 
\hline
\label{tab2}
\end{tabular}
\end{minipage}
\end{table}

Our interferometric observations were made with the PAVO beam combiner \citep{Ireland08} at the CHARA Array at Mt. 
Wilson Observatory, California \citep{tenBrummelaar05}. PAVO is a pupil-plane beam combiner, optimised for high 
sensitivity (limiting magnitude in typical seeing conditions of $R\sim8$) at visible wavelengths ($\sim$600 to 900\,nm).
Two or three beams may be combined. Through spectral dispersion each scan typically produces visibility measurements 
in 20 independent wavelength channels. 
With available baselines up to 330\,m, PAVO at CHARA is one of the highest angular-resolution instruments operating 
worldwide. Further details on this instrument were given by \citet{Ireland08}. Early PAVO science results have been
presented by \citet{Bazot11}, \citet{Derekas11}, \citet{Huber12b,Huber12a} and \citet{Maestro12}. 

Most of our observations were made during several nights in August 2012, although some data were taken
during previous observing seasons in 2010 and 2011. Our observations have been made using PAVO in two-telescope mode,
with baselines ranging from 110~to~250\,m. A summary of our observations is given in Table~\ref{tab1}. To
calibrate the fringe visibilities in our targets we observed nearby stars, which ideally would be
unresolved point sources with no close companions. In practice we used stars as unresolved as possible, which in our case meant spectral
types A and B. Table~\ref{tab2} lists the calibrators used in our analysis. We determined the expected angular diameters 
of the calibrators using the ($V-K$) relation of \citet{Kervella04}. We adopted $V$-band magnitudes from the Tycho 
catalogue \citep{Perryman97} and converted them into the Johnson system using the calibration by \citet{Bessell00}. 
$K$-band magnitudes were taken from the Two Micron All Sky Survey \citep[2MASS;][]{Skrutskie06}. To de-redden the 
photometry we used the extinction model of \citet{Drimmel03} to estimate interstellar reddening, and adopted the 
reddening law of \citet{ODonnell94} \citep[see also][]{Cardelli89}.

\begin{figure*}
\center
\includegraphics[scale=1.5]{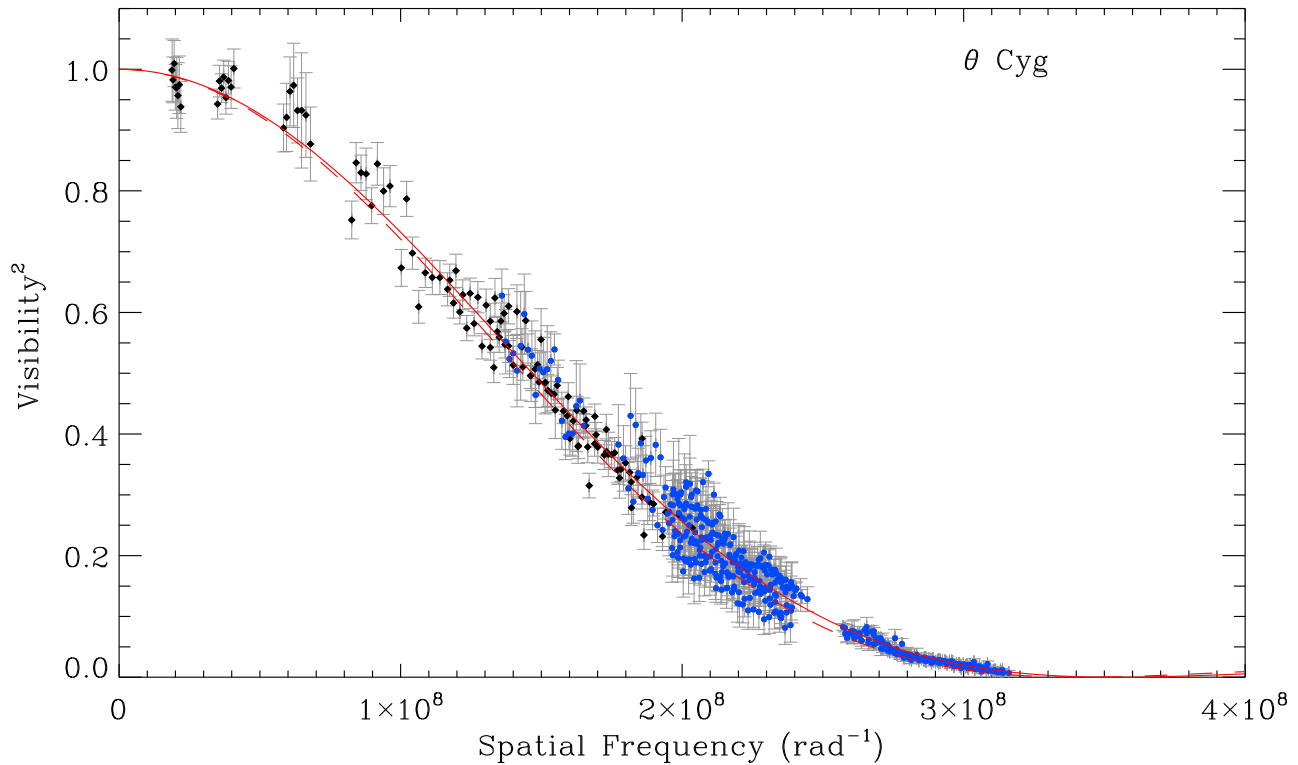}
\caption{Squared visibility versus spatial frequency for $\theta$ Cyg for PAVO (blue circles) and MIRC (black diamonds) 
data. The red lines show the fitted limb-darkened model to the combined data. The solid line is 
for $\mu$=0.47$\pm$0.04 (PAVO) while the dashed line is for $\mu$=0.21$\pm$0.03 (MIRC).}
\label{fig1}
\end{figure*}

In the case of our largest calibrator, HD\,188665, there is some indication that it is larger than expected from 
the ($V-K$) relation ($\theta_{V-K}=0.240$\,mas). Calibrating with smaller calibrator stars observed at similar times
we find the average interferometric response of HD\,188665 is consistent with a uniform-disc diameter of 
$\theta_{UD}=0.274\pm0.008$\,mas.
 
To best account for temporal variations in system visibility due to changes in seeing, calibrators must be observed as closely spaced in time to 
the targets as possible. We observed $\theta$~Cyg and its calibrators in the sequence: {\it calibrator 1 -- $\theta$~Cyg 
-- calibrator 2}, with a scan of each object obtaining two minutes of visibility data. Such a sequence typically lasted 
15\,minutes, including slewing. To minimise slew times when observing 16~Cyg~A~and~B, we observed in the sequence: 
{\it calibrator 1 -- 16~Cyg~A -- 16~Cyg~B -- calibrator 2}. This typically took 20\,minutes.

In addition to our observations with PAVO, supporting observations were also made in the infrared with two other
beam combiners at the CHARA Array. Observations of $\theta$~Cyg were made using the Michigan Infrared
Combiner \citep[MIRC;][]{Monnier04}, while observations of 16~Cyg~A~and~B were made with the CHARA Classic combiner.

MIRC combines up to six telescope beams in the image-plane,
allowing for simultaneous visibility measurements on 15 baselines and 20 closure phase measurements. Additionally, MIRC
splits the $H$-band light ($\lambda_0=1.65\mu$m) into eight independent spectral channels. Further details on the MIRC
instrument may be found in \citet{Monnier04, Monnier06, Monnier10} and \citet{Che10, Che12}. Our MIRC observations 
consist of four scans of $\theta$~Cyg, made in six-telescope mode on 19 June 2012. 
The calibrator star used was $\sigma$~Cyg, with an assumed diameter of $\theta_{UD}=0.54\pm0.02$\,mas \citep{Barnes78}.

Classic is a pupil-plane combiner operating in a two-telescope mode in either of $H$-band ($\lambda_0=1.65\mu$m) 
or $K^{\prime}$-band ($\lambda_0=2.14\mu$m). The Classic observations of 16~Cyg~A~and~B were previously presented by 
\citet{Boyajian13}. Observations were made in $H$-band, with 23 and 24 brackets of 16~Cyg~A~and~B, 
respectively, over the nights of 16, 19, 20 and 21 August 2011 using the S1E1, E1W1 and S1W1 baselines. Calibration 
stars were HD\,185414, HD\,191096 and HD\,191195, whose estimated angular diameters were taken from the SearchCal tool 
developed by the JMMC Working Group \citep{Bonneau06, Bonneau11}.

For each target we fitted a limb-darkened disc model to the visibility measurements \citep{HanburyBrown74},
\begin{equation*}
V = \left( \frac{1-\mu_\lambda}{2} + \frac{\mu_\lambda}{3} \right)^{-1} \qquad\qquad\qquad\qquad\qquad\qquad
\end{equation*}
\begin{equation}
  \qquad{} \times \left[ (1-\mu_\lambda) \frac{J_1(x)}{x} + \mu_\lambda (\pi/2)^{1/2} \frac{J_{3/2}(x)}{x^{3/2}} \right],\label{eqn1}
\end{equation}
where
\begin{equation}
x = \pi B \theta_\mathrm{LD} \lambda^{-1}.\label{eqn2}
\end{equation}
Here, $V$ is the visibility, $\mu_\lambda$ is the linear limb-darkening coefficient, $J_n(x)$ is the $n^\mathrm{th}$ 
order Bessel function, $B$ is the projected baseline, $\theta_\mathrm{LD}$ is the angular diameter after correction for
limb-darkening, and $\lambda$ is the wavelength at which the observations was made. The quantity $B\lambda^{-1}$ is
often referred to as the spatial frequency.

We determined linear limb-darkening coefficients in the $R$ and $H$ bands for our targets by interpolating the model 
grid by \citet{Claret11} to the spectroscopic estimates of $T_\mathrm{eff}$, log $g$ and [Fe/H] given in Table~\ref{tab0}
for a microturbulent velocity of 2~km~s$^{-1}$. The uncertainties in the spectroscopic parameters were used to create 1000
realisations of the limb-darkening coefficients, from which the uncertainties were estimated. Adopted values of the linear 
limb-darkening coefficients are given in Table~\ref{tab3}. Typically the influence of the adopted limb-darkening on the 
final fitted angular diameter is relatively small. Detailed 3-D hydrodynamical models by \citet{Bigot06}, \citet{Chiavassa10} 
and \citet{Chiavassa12} for dwarfs and giants have shown that the differences from simple linear limb-darkening models are 
$\sim$1\% or less in angular diameter for stars with near-solar metallicity. For a moderately-well resolved star with 
$V^2 \sim 0.5$, a 1\% change in angular diameter would correspond to an uncertainty of less than 1\% in $V^2$. For
our measurements these effects may be non negligible and our results will be valuable for comparing simple 1-D to 
sophisticated 3-D models.

To fit the model and estimate the uncertainty in the derived angular diameters we followed the procedure outlined by
\citet{Derekas11}. This involved performing Monte Carlo simulations, taking into account uncertainties in the data,
adopted wavelength calibration (0.5\% for PAVO, 0.25\% for MIRC), calibrator sizes (5\%) and limb-darkening 
coefficients (see Table~\ref{tab3}).

\begin{table*}
 \centering
 \begin{minipage}{140mm}
  \caption{Measured Angular Diameters and Fundamental Properties}
  \begin{tabular}{@{}lcccccccccc@{}}
  \hline
   Star & Combiner & $\mu$ & $\theta_\mathrm{UD}$ (mas) & $\theta_\mathrm{LD}$ (mas) & $R$ (R$_\odot$) & $M$ (M$_\odot$) & $T_\mathrm{eff}$ (K) \\
 \hline
$\theta$ Cyg & PAVO               & 0.47$\pm$0.04 & 0.720$\pm$0.004 &      0.754$\pm$0.009  &      1.49$\pm$0.02  &      1.37$\pm$0.04  &      6745$\pm$44\\
             & MIRC               & 0.21$\pm$0.03 & 0.726$\pm$0.014 &      0.739$\pm$0.015  &      1.46$\pm$0.03  &      1.31$\pm$0.06  &      6813$\pm$72\\
             & {\bf PAVO+MIRC}    &      ...      &       ...       & {\bf 0.753$\pm$0.009} & {\bf 1.48$\pm$0.02} & {\bf 1.37$\pm$0.04} & {\bf 6749$\pm$44}\\
16 Cyg A     & PAVO               & 0.54$\pm$0.04 & 0.513$\pm$0.004 &      0.539$\pm$0.006  &      1.22$\pm$0.02  &      1.07$\pm$0.04  &      5839$\pm$37\\
             & Classic            & 0.26$\pm$0.04 & 0.542$\pm$0.015 &      0.554$\pm$0.016  &      1.26$\pm$0.04  &      1.16$\pm$0.10  &      5759$\pm$85\\
             & {\bf PAVO+Classic} &      ...      &       ...       & {\bf 0.539$\pm$0.007} & {\bf 1.22$\pm$0.02} & {\bf 1.07$\pm$0.05} & {\bf 5839$\pm$42}\\
16 Cyg B     & PAVO               & 0.56$\pm$0.04 & 0.467$\pm$0.004 &      0.490$\pm$0.006  &      1.12$\pm$0.02  &      1.05$\pm$0.04  &      5809$\pm$39\\
             & Classic            & 0.27$\pm$0.04 & 0.502$\pm$0.020 &      0.513$\pm$0.020  &      1.17$\pm$0.05  &      1.20$\pm$0.14  &      5680$\pm$112\\
             & {\bf PAVO+Classic} &      ...      &       ...       & {\bf 0.490$\pm$0.006} & {\bf 1.12$\pm$0.02} & {\bf 1.05$\pm$0.04} & {\bf 5809$\pm$39}\\
\hline
\label{tab3}
\end{tabular}
\end{minipage}
\end{table*}

\section{Results}

\subsection{Fundamental stellar properties}\label{4_1}
Combining our interferometric measurements with astrometric, asteroseismic and photometric measurements allows us
to derive radii, masses and effective temperatures that are nearly model-independent.

The linear radius, $R$, is,
\begin{equation}
R=\frac{1}{2}\theta_\mathrm{LD}D,\label{eqn3}
\end{equation}
where $D$ is the distance to the star, which itself is obtained directly from the parallax.

From an estimate of the bolometric flux at Earth, $F_\mathrm{bol}$, we can find the effective temperature,
\begin{equation}
T_\mathrm{eff}=\left(\frac{4F_\mathrm{bol}}{\sigma\theta_\mathrm{LD}^2}\right)^{1/4},\label{eqn4}
\end{equation}
where $\sigma$ is the Stefan-Boltzmann constant.

Finally, to obtain the mass we use the scaling relation between the large frequency separation of solar-like oscillations,
$\Delta\nu$, and the density of the star \citep{Ulrich86}:
\begin{equation}
\frac{\Delta\nu}{\Delta\nu_\odot}=\left(\frac{M}{\mathrm{M}_\odot}\right)^{1/2}\left(\frac{R}{\mathrm{R}_\odot}\right)^{-3/2}.\label{eqn5}
\end{equation}
It follows from this relation that we can derive the mass of the star from measurements of the angular diameter,
parallax and large frequency separation. Caution is required when using this relationship because the assumption that 
leads to this relation, that other stars are homologous to the Sun, is not strictly valid \citep{Belkacem12}, although
it has been shown that the relation holds to within 5\% in models \citep{Stello09,White11a}. Particular care is needed
for stars above ~1.2\,M$_\odot$ and beyond the main-sequence, since models indicate a departure that is largely a 
function of effective temperature \citep{White11a}. We note that one should measure $\Delta\nu$ of the stars and the
Sun in a self-consistent manner. We adopt a solar value of $\Delta\nu_\odot$\,=\,135.1\,$\mu$Hz.

\subsection{$\theta$ Cyg}
\begin{figure}
\center
\includegraphics[scale=0.8]{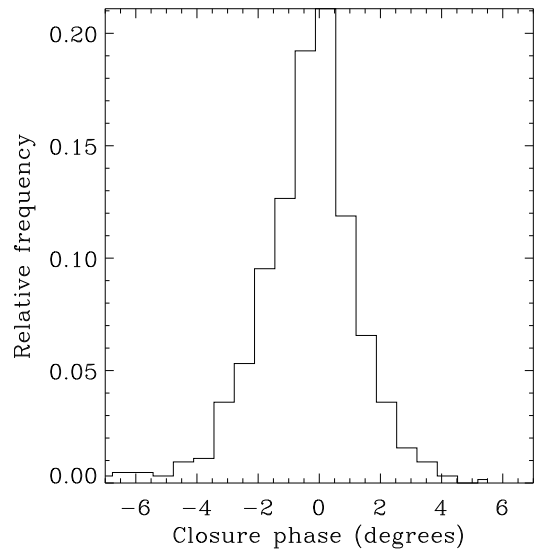}
\caption{Histogram of MIRC closure phase measurements for $\theta$ Cyg.}
\label{fig2}
\end{figure}

Figure \ref{fig1} presents the calibrated squared-visibility measurements as a function of spatial frequency for
$\theta$ Cyg. We have performed limb-darkened fits to the PAVO and MIRC data both separately and together.
For the combined fit we fitted a common angular diameter, but applied a different linear limb-darkening coefficient to 
the MIRC and PAVO data. Provided the star has a compact atmosphere, the limb-darkened angular
diameter should be independent of wavelength. We also fitted uniform-disc models separately. Uniform-disc diameters are 
wavelength dependent due to the effects of limb darkening. 

The fitted angular diameters are given in Table~\ref{tab3}, along with the radius, mass and effective temperature
(see Sec.~\ref{4_1}). We note that our diameter ($\theta_\mathrm{LD}=0.753\pm0.009$\,mas) is consistent with that obtained 
recently by \citet[][$\theta_\mathrm{LD}=0.760\pm0.003$\,mas]{Ligi12}, as well as the estimation by 
\citet[][$\theta_\mathrm{LD}=0.760\pm0.021$\,mas]{vanBelle08}. The values found by 
\citet[][$\theta_\mathrm{UD}=0.845\pm0.015$\,mas and $\theta_\mathrm{LD}=0.861\pm0.015$\,mas]{Boyajian12a} 
are inconsistent with our data. Operating at higher spatial frequencies, PAVO is better able to resolve $\theta$ Cyg than 
Classic. With the lower resolution of Classic, calibration errors have greater impact and this could explain the 
discrepancy.

We note that the uncertainty in the PAVO diameter is dominated by our adopted uncertainties in the limb-darkening 
coefficient rather than measurement uncertainties. Uncertainties in the calibrator sizes and wavelength scale also make
significant contributions to the overall error budget. For comparison, ignoring these uncertainties in fitting the PAVO 
data yields a fractional uncertainty of only 0.2\% compared to an uncertainty of 1.2\% derived from our Monte Carlo 
simulations. This illustrates the importance of taking into account these additional uncertainties.

To determine the mass we have used the revised scaling relation for $\Delta\nu$ proposed by \citet{White11a}, which 
corrects for a deviation from the original scaling relation that is dependent upon effective temperature. Without 
this correction we obtain a significantly lower mass for $\theta$ Cyg (1.27\,M$_\odot$) that is not consistent with 
the value of 1.39$^{+0.02}_{-0.01}$\,M$_\odot$ obtained from isochrones by \citet{Casagrande11} in their re-analysis 
of the Geneva-Copenhagen Survey \citep{Nordstrom04, Holmberg07, Holmberg09}.

For calculating the effective temperature we have used the bolometric flux determined from spectrophotometry by 
\citet{Boyajian13} (see Table~\ref{tab0}). In addition to the formal errors quoted by \citet{Boyajian13}, we 
include an additional 1\% uncertainty accounting for systematics present in the absolute flux calibration of 
photometric data \citep[see discusion in][]{Bessell12}.
Our measured temperature for $\theta$ Cyg (6749$\pm$44\,K) is in excellent
agreement with the values determined by \citet{Erspamer03} from spectroscopy (6745$\pm$150\,K) and \citet{Ligi12} from
interferometry (6767$\pm$87\,K).

Figure ~\ref{fig2} shows a histogram of the MIRC closure phase measurements. All values are consistent with zero, 
indicating the source has a point-symmetric intensity distribution. \citet{Ligi12} reported that the scatter in their
measurements of $\theta$~Cyg with the VEGA beam combiner was higher than expected, leaving open the possibility of 
stellar variations or a close companion. As they noted, {\it Kepler} observations would have detected any large stellar 
pulsations, and so this explanation for their result is unsatisfactory. 

Our MIRC closure phase measurements also appear to rule out a new close companion. Following the method used by 
\citet{Kraus08} we estimated the detection threshold for a close companion as a function of separation. Briefly, 
this method involves Monte Carlo simulations of data sets with the same ($u$, $v$)-sampling and error properties of our MIRC 
observations and finding the best-fit contrast ratio within a large grid of positions and separations. The 
99.9\% upper limit to companion brightness within a series of annuli was determined as the contrast ratio for 
which 99.9\% of simulations had no companion brighter than this limit anywhere within the annulus. We find that a potential 
close companion with a separation between 10--20\,mas ($\sim$0.2--0.4\,AU) must be at least 4.68\,mag fainter in $H$-band than 
$\theta$ Cyg to escape detection in our observations. For separations between 20--40\,mas ($\sim$0.4--0.7\,AU), the companion 
must be at least 3.44\,mag fainter.

\subsection{16 Cyg A and B}

\begin{figure}
\center
\includegraphics[scale=0.68]{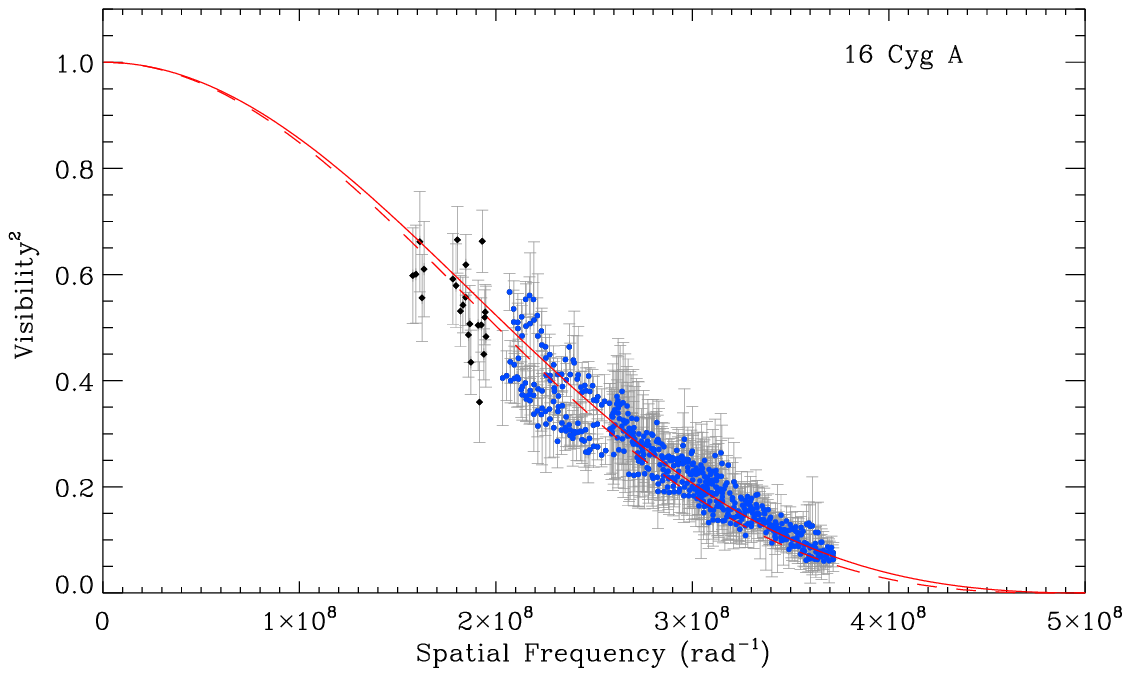}
\includegraphics[scale=0.68]{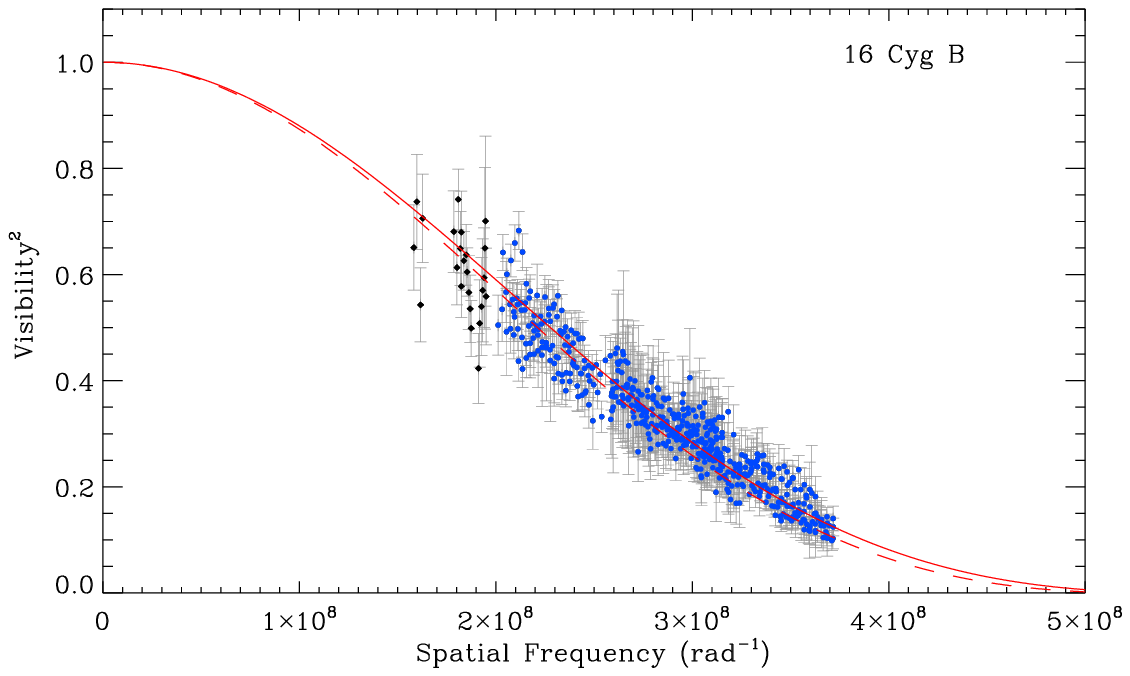}
\caption{Squared visibility versus spatial frequency for 16 Cyg A (top) and B (bottom) for PAVO (blue circles)
and Classic (black diamonds) data. The red lines show the fitted limb-darkened model to the combined data. The solid 
lines use the limb-darkening coefficients in $R$-band (PAVO) while the dashed line is for $H$-band (Classic).
Note that the error bars for each star have been scaled so that the reduced $\chi^2$ equals unity.}
\label{fig3}
\end{figure}

Figure \ref{fig3} shows the calibrated squared-visibility PAVO and Classic measurements as a function of spatial 
frequency for 16 Cyg A and B. As for $\theta$ Cyg, we provide the fitted uniform-disc and limb-darkened diameters, 
along with derived properties in Table~\ref{tab3}. Inclusion of the Classic data in the fit does not significantly 
change the measured diameters. 

The diameters as measured individually with PAVO and Classic agree within the uncertainties. Our 16~Cyg~B 
measurement is 1.1$\sigma$ larger than the diameter measured by 
\citet[][$\theta_\mathrm{LD}$\,=0.426\,$\pm$\,0.056\,mas]{Baines08}, although their estimate from spectral energy 
distribution fitting ($\theta_\mathrm{LD}$\,=0.494\,$\pm$\,0.019\,mas) is in excellent agreement with our final 
value. 

The larger uncertainties in our Classic diameters compared to the values reported by \citet{Boyajian13} arise 
largely from the inclusion of additional uncertainty in the linear limb-darkening coefficient. It is also worth noting 
that whereas we determined the limb-darkening coefficient from the model grid of \citet{Claret11}, \citet{Boyajian13} used 
the values from \citet{Claret00}, leading to slightly different values being used. 

We are able to compare our measured radii and masses with those obtained by preliminary asteroseismic modelling by 
\citet{Metcalfe12}. Several different approaches were taken to model the pair, using measured oscillation frequencies
and spectroscopic values as constraints. Methods varied with different stellar evolutionary and pulsation codes, 
nuclear reaction rates, opacities, and treatments of diffusion and convection used. 

As mentioned in Section~\ref{16Cyglit}, the best fit model of each method fell into one of two families. A low radius, 
low mass family favoured $R$\,=\,1.24\,$\mathrm{R_\odot}$, $M$\,=\,1.10\,$\mathrm{M_\odot}$ for 16~Cyg~A and 
$R$\,=\,1.12\,$\mathrm{R_\odot}$, $M$\,=\,1.05\,$\mathrm{M_\odot}$ for 16~Cyg~B. The high radius, high mass family 
favoured $R$\,=\,1.26\,$\mathrm{R_\odot}$, $M$\,=\,1.14\,$\mathrm{M_\odot}$ and $R$\,=\,1.14\,$\mathrm{R_\odot}$, 
$M$\,=\,1.09\,$\mathrm{M_\odot}$, respectively.

Comparison with our results in Table~\ref{tab3} shows a preference for the low radius, low mass family, although
the high radius, high mass family cannot be completely discounted, particularly for 16~Cyg~B. This brief comparison 
suggests that, in conjunction with more {\it Kepler} data that is becoming available, our interferometric results
will help to significantly constrain stellar models.

When calculating the effective temperature we have again used the bolometric flux determined by \citet{Boyajian13},
once again adopting an additional 1\% uncertainty to account for systematics in the absolute flux calibration. 
As for $\theta$ Cyg, our measured temperatures (5839$\pm$42\,K and 5809$\pm$39\,K for
16~Cyg~A~and~B, respectively) agree well with the spectroscopically determined values 
\citep[5825$\pm$50\,K and 5750$\pm$50\,K;][]{Ramirez09}.

\subsection{Comparison with asteroseismic scaling relations}

In addition to the scaling relation for the large frequency separation, $\Delta\nu$, given in Equation~\ref{eqn5},
there is also a widely used scaling relation for the frequency of maximum power, $\nu_\mathrm{max}$ 
\citep{Brown91,Kjeldsen95}:
\begin{equation}
\frac{\nu_\mathrm{max}}{\nu_\mathrm{max,\odot}}=\left(\frac{M}{\mathrm{M}_\odot}\right)\left(\frac{R}{\mathrm{R}_\odot}\right)^{-2}\left(\frac{T_\mathrm{eff}}{\mathrm{T_{eff,\odot}}}\right)^{-1/2}.\label{eqn6}
\end{equation}
Equations~\ref{eqn5}~and~\ref{eqn6} may be simultaneously solved for mass and radius:

\begin{equation}
\frac{M}{\mathrm{M}_\odot}=\left(\frac{\nu_\mathrm{max}}{\nu_\mathrm{max,\odot}}\right)^{3}\left(\frac{\Delta\nu}{\Delta\nu_\odot}\right)^{-4}\left(\frac{T_\mathrm{eff}}{\mathrm{T_{eff,\odot}}}\right)^{3/2}\label{eqn7}
\end{equation}
and
\begin{equation}
\frac{R}{\mathrm{R}_\odot}=\left(\frac{\nu_\mathrm{max}}{\nu_\mathrm{max,\odot}}\right)\left(\frac{\Delta\nu}{\Delta\nu_\odot}\right)^{-2}\left(\frac{T_\mathrm{eff}}{\mathrm{T_{eff,\odot}}}\right)^{1/2}.\label{eqn8}
\end{equation}

Provided the effective temperature is known, the stellar mass and radius may be estimated directly from the asteroseismic 
parameters $\Delta\nu$ and $\nu_\mathrm{max}$. This is sometimes referred to as the `direct method' 
\citep{Kallinger10d,Chaplin11a,SilvaAguirre11}
in contrast to determining mass, radius and other parameters via stellar modelling \citep{Stello09b,Basu10,Kallinger10b,Gai11}.

The scaling relation for $\Delta\nu$, which we have used to derive the masses in Table~\ref{tab3}, is better understood
theoretically with tests of its validity in models finding the relation holds to within 5\% \citep{Stello09,White11a}.
The $\nu_\mathrm{max}$ scaling relation relies on the argument that $\nu_\mathrm{max}$ should scale with the
acoustic cutoff frequency \citep{Brown91}, although the underlying physical reason for this relationship has not been
clear. Only recently has the theoretical framework behind this result begun to be developed \citep{Belkacem11}.
Understanding the validity of these scaling relations has become particularly important as they are now commonly used
to determine radii for a large number of faint {\it Kepler} stars, including some stars with detected exoplanet
candidates \citep[see, e.g.,][]{Borucki12,Huber13}. We are able to test the validity of the asteroseismic scaling 
relations by comparing our interferometric radii with independently determined asteroseismic radii calculated using 
Equation~\ref{eqn8}. To ensure the asteroseismic radii are truly independent of our interferometric radii,
in this calculation we use the spectroscopic effective temperatures given in Table~\ref{tab0}.

We have determined the global asteroseismic properties, $\Delta\nu$ and $\nu_\mathrm{max}$, of 16~Cyg~A~and~B 
using the automated analysis pipeline by \citet{Huber09}, which has been shown to agree well with other methods 
\citep{Hekker11,Verner11}. These values are given in Table~\ref{tab4}, along with the radii derived from the scaling
relations, Equations~\ref{eqn5}~and~\ref{eqn6}. We use solar values of $\Delta\nu_\odot$\,=\,135.1\,$\mu$Hz and
$\nu_\mathrm{max,\odot}$\,=\,3090\,$\mu$Hz.

We do not consider $\theta$ Cyg here because the width of the
oscillation envelope is very broad, which makes $\nu_\mathrm{max}$ ambiguous. This, along with large mode linewidths
\citep{Chaplin09,Baudin11,Appourchaux12a, Corsaro12}, appears to be a feature of oscillations in F stars. Observations
of the F~subgiant Procyon showed a similarly broad envelope \citep{Arentoft08}.

\begin{figure}
\center
\includegraphics[scale=0.4]{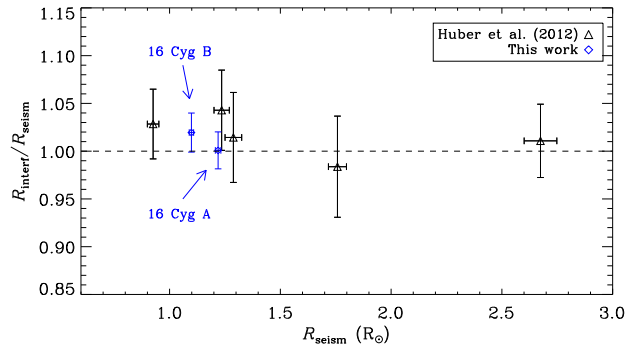}
\caption{Comparison of stellar radii measured using interferometry and calculated using asteroseismic scaling relations.
Black triangles show stars measured by \citet{Huber12b}, while blue diamonds show 16~Cyg~A~and~B.}
\label{fig4}
\end{figure}

Figure~\ref{fig4} shows the remarkable agreement between the interferometric and asteroseismic radii. In addition to 
16~Cyg~A~and~B we also include five stars for which \citet{Huber12b} determined interferometric and asteroseismic radii using the
same method (see their Figure~7). The agreement for 16~Cyg~A~and~B is within 1$\sigma$ and at a $\sim$2\% level,
which makes this the most precise independent empirical test of asteroseismic scaling relations yet. However, further
studies are still needed, particularly of stars that are significantly different from the Sun, to robustly test the
validity of the scaling relations.

\begin{table}
 \centering
 \begin{minipage}{140mm}
  \caption{Asteroseismic Properties and Radii of 16 Cyg A and B}
  \begin{tabular}{@{}lcc@{}}
  \hline
    & 16 Cyg A & 16 Cyg B\\
 \hline
$\Delta\nu$ ($\mu$Hz) & 103.5$\pm$0.1 & 117.0$\pm$0.1\\
$\nu_\mathrm{max}$ ($\mu$Hz) & 2201$\pm$20 & 2552$\pm$20\\
$R$ (R$_\odot$) & 1.218$\pm$0.012 & 1.098$\pm$0.010\\
\hline
\label{tab4}
\end{tabular}
\end{minipage}
\end{table}

\section{Conclusions}
We have used long-baseline interferometry to measure angular diameters for $\theta$ Cyg and 16 Cyg A and B. 
All three stars have been observed by the {\it Kepler Mission} and exhibit solar-like oscillations, allowing for
detailed study of their internal structure.

For $\theta$ Cyg we find a limb-darkened angular diameter of $\theta_\mathrm{LD}=0.753\pm0.009$ mas, which, combined
with the Hipparcos parallax, gives a linear radius of $R$\,=\,1.48$\pm$0.02\,$\mathrm{R_\odot}$. When determining the
mass (1.37$\pm$0.04\,$\mathrm{M_\odot}$) from the interferometric radius and large frequency separation, $\Delta\nu$, 
we find that it is necessary to use the revised scaling relation for $\Delta\nu$ suggested by \citet{White11a}. 
This revision takes into account a deviation from the standard scaling relation in stars of higher temperature, without 
which the determined mass would be significantly lower (1.27\,$\mathrm{M_\odot}$) than expected from fitting to 
isochrones.

Closure phase measurements of $\theta$ Cyg reveal the star to be point symmetric, consistent with being a single
star. This rules out the possibility of all but a very low luminosity close companion, which had previously been 
suggested.

For 16~Cyg~A~and~B we have found limb-darkened angular diameters of $\theta_\mathrm{LD}=0.539\pm0.007$ and 
$\theta_\mathrm{LD}=0.490\pm0.006$, and linear radii of $R$\,=\,1.22$\pm$0.02\,$\mathrm{R_\odot}$ 
and $R$\,=\,1.12$\pm$0.02\,$\mathrm{R_\odot}$, respectively. Comparing these radii with those derived from
the asteroseismic scaling relations shows good agreement at a $\sim$2\% level. 

Our measurements of near-model-independent masses, radii and effective temperatures will provide
strong constraints when modelling these stars.

\section*{Acknowledgments}
The CHARA Array is funded by the National Science Foundation through
NSF grant AST-0606958, by Georgia State University through the College of Arts and Sciences, and by the W.M. Keck
Foundation. We acknowledge the support of the Australian Research Council. Funding for the Stellar Astrophysics Centre
is provided by The Danish National Research Foundation. We acknowledge the {\it Kepler} Science Team and all those who 
have contributed to the {\it Kepler Mission}. Funding for the {\it Kepler Mission} is provided by NASA's Science 
Mission Directorate. T.R.W. is supported by an Australian Postgraduate Award, a University of Sydney Merit Award, an 
Australian Astronomical Observatory PhD Scholarship and a Denison Merit Award. DH is supported by an appointment to the 
NASA Postdoctoral Program at Ames Research Center, administered by Oak Ridge Associated Universities through a contract 
with NASA.

\label{lastpage}

\end{document}